\documentclass[12pt]{iopart}
\usepackage{graphicx}
\usepackage{bm}
\usepackage{amssymb}

\begin{document}
\title[
Josephson vortex motion as a source for dissipation of superflow
of e-h pairs]{Josephson vortex motion as a source for dissipation
of superflow of e-h pairs in  bilayers}
\author{D V Fil$^1$ and S I Shevchenko$^2$}
\address{$^1$Institute for Single Crystals, National Academy of
Science of Ukraine, Lenin av. 60, Kharkov 61001, Ukraine\\
$^2$B. Verkin Institute for Low Temperature Physics and
Engineering, National Academy of Sciences of Ukraine, Lenin av. 47
Kharkov 61103, Ukraine} \eads{
\mailto{fil@isc.kharkov.ua},\mailto{shevchenko@ilt.kharkov.ua}}

\begin{abstract}
It is shown that  in a bilayer excitonic superconductor
dissipative losses emerge under transmission of the current from
the source to the load. These losses are proportional to the
square of the interlayer tunneling amplitude and independent on
the value of the input current. The case of quantum Hall bilayer
is considered. The bilayer may work as a transmission line if the
input current exceeds certain critical value. The input current
higher than critical one induces Josephson vortices in the
bilayer. The difference of electrochemical potentials is required
to feed the load and it forces Josephson vortices to move. The
state becomes non-stationary that leads to dissipation.
\end{abstract}

\pacs{73.43.Jn,   74.90.+n}


\section{Introduction}
\label{intro}

Among the phenomena that  demonstrate bilayer electron systems in
semiconductor heterostructures considerable attention is given to
superfluidity of electrons-hole (e-h) pairs with components
belonging to different layers (see, for instance, \cite{em}).
 A flow of electron-hole pairs
in the bilayer is equivalent to two oppositely directed electrical
currents in  the layers. Therefore, the superflow of such pairs is
a kind of superconductivity. For the first time the effect was
considered in Refs. \cite{1,2} for bilayers where one layer is of
the electron-type conductivity and the other layer -  of the
hole-type one (electron-hole bilayers). The next important step
was the prediction of the e-h superfluidity for the systems where
the conductivity of both layers is of the same type \cite{4,5}. In
that case the bilayer should be subjected by a strong
perpendicular to the layers magnetic field. If the total filling
factor of the Landau levels is $\nu=1$ the number of electrons in
one layer coincides with the number of holes  in the other layer
(the empty states in the lowest Landau level play the role of
holes) and the Coulomb attraction between electrons and holes
results in a formation of bound pairs. Note that the description
of electron-hole pairing in electron-electron bilayers in a
quantized magnetic field is close to one developed earlier for the
quantum Hall electron-hole bilayers \cite{kh, ll, rs}

The prediction \cite{4,5} has inspired considerable increase of
interest to the study of this phenomenon, both theoretically
\cite{t1,t2,t3,sh1,sh2, lr,18,t7,t8} and
experimentally\cite{9,12,di,6,7,8,e9,e10,e11}. The results of
recent experimental investigations of quantum Hall bilayers
support the idea on  superfluidity of e-h pairs in these systems.
In particular, in the counterflow experiments a huge increase of a
longitudinal conductivity was observed\cite{9,12,di}. Other
important observations are a large low bias tunnel conductivity
\cite{6}, that also takes place in bilayers with large imbalance
of filling factors \cite{e11}, the Goldstone collective mode
\cite{7},  the quantized Hall drag between the layers \cite{8},
the interlayer drag \cite{e9} and the interlayer critical
supercurrent \cite{e10} in the Corbino disk geometry. The low
temperature properties of optically generated indirect excitons in
bilayers in zero magnetic field were also studied experimentally
\cite{13,14}, and specific features in the photoluminescence
spectra accounted for the Bose-Einstein condensation of
electron-hole pairs have been observed. Recent important
contribution into this topic is connected with the idea of using
two graphene layers separated by a dielectric layer
\cite{g1,g2,g3,g4} instead of GaAs heterostructures. It is
expected \cite{g1} that in graphene systems in zero magnetic field
superconductivity of e-h pairs may take place at rather high
temperatures.

A superfluid state of electron-hole pairs in quantum Hall bilayers
can be considered as a state with spontaneous interlayer phase
coherence between the electrons. The coherent phase $\varphi$ is
the phase of the order parameter for the electron-hole pairing,
and  gradient of the phase $\varphi$ determines the value of the
supercurrent in the layer (planar supercurrent).

The bilayer system in the counterflow setup\cite{9,12,di} can be
used as a double-wire circuit that transmits the current from the
source to the load. If the transmission is provided by superfluid
e-h pairs one can expect that such a line will work as a usual
superconducting transmission line. Nevertheless, genuine
superconductivity was not achieved in the counterflow experiments.
A finite resistance can be accounted for a disorder that results
in a finite concentration of planar vortices in the bilayer
\cite{d1,d2}. But there is another principal circumstance that may
forbid genuine superconductivity in such a transmission line. To
support nonzero current in the load circuit the difference of
electrochemical potentials between the layers is required. This
difference results in a temporal dependence of the phase
$\varphi$. The state with the spontaneous interlayer phase
coherence   emerges if  the layers are situated rather close to
each other (at the distance less or of other of the magnetic
length). Therefore, nonzero interlayer tunneling is always present
in real physical systems. At nonzero interlayer tunneling
amplitude  the difference of electrochemical potentials results in
appearance of an a.c. Josephson current and the state becomes
non-stationary one. As was shown in \cite{j1} dissipative losses
emerge in the non-stationary state. The situation is different for
the loop and for the load geometry \cite{j2}. The steady state is
not possible in the loop geometry Fig. {\ref{f1}b (that
corresponds to the transmission line \cite{9,12,di}) but it can be
realized in the load geometry shown in Fig. \ref{f1}a under
condition that the difference of the electrochemical potential of
the layers is tuned to zero.

\begin{figure}
\begin{center}
\includegraphics[width=10cm]{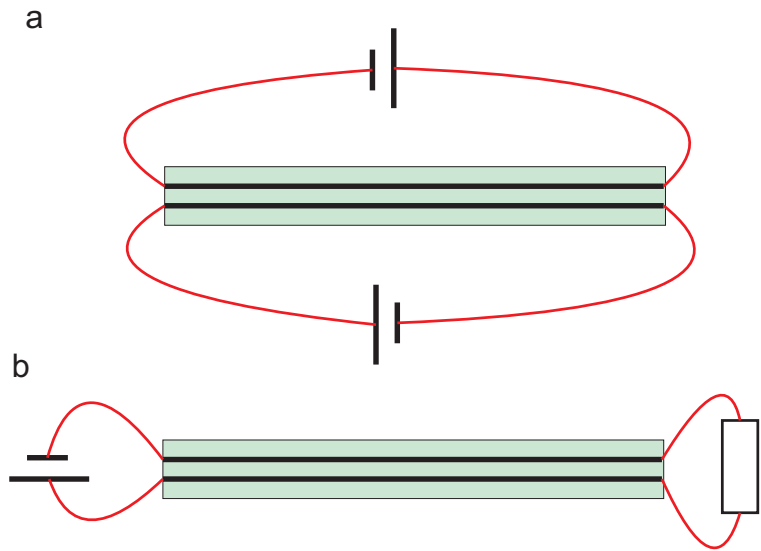}
\end{center}
\caption{The  load (a) and the loop (b) setups.} \label{f1}
\end{figure}

The load geometry \cite{j2} is an appropriate setup for the
observation of the superfluidity of e-h pairs but it cannot be
used for the transmission of the current from the source to the
load. In more complicated circuits, e.g. made of a stack of
bilayers, non-dissipative transmission of the energy is not
possible, as well. Indeed, a bilayer with zero difference of
electrochemical potentials can be shunted at both ends without any
impact on electrical currents and voltages in external circuits.
It means that the removal of such a bilayer from the circuit
cannot change its working parameters, including the power of
losses. An example of such shunting is shown in Fig. \ref{f2}. The
only setup where superfluid properties of e-h pairs may lead to
lowering of dissipation is the system with nonzero difference of
electrochemical potentials between the layers.  While
non-stationary states are not free from dissipation, the losses
can be exponentially small and one can say about almost
superconducting state.

\begin{figure}
\begin{center}
\includegraphics[width=10cm]{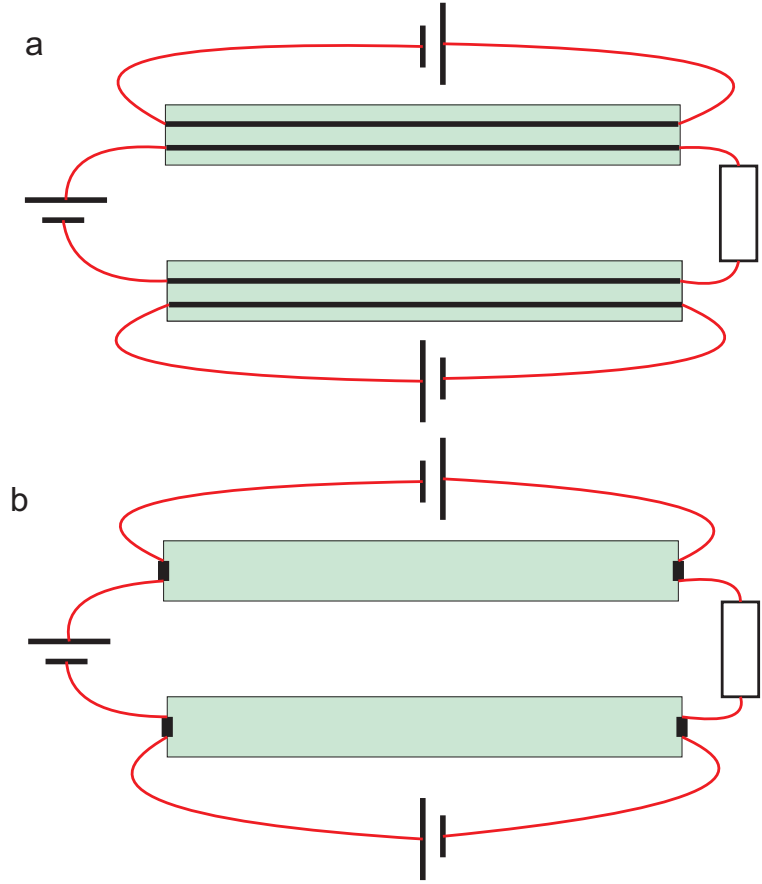}
\end{center}
\caption{A circuit with a stack of bilayers (a) and an equivalent
circuit with shunted bilayers (b)} \label{f2}
\end{figure}

In this paper we consider dissipative processes in the bilayer in
the loop geometry and find the conditions at which the losses are
small.  The state with a.c. Josephson and planar supercurrents
that emerges in the bilayer can be described as a moving chain of
Josephson vortices. Under such a motion two mechanisms of
dissipation come into play. They are the second viscosity that
results in dissipation in the non-stationary regime and Joule
losses caused by a.c. electrical fields that emerge due to spatial
and temporal variation of the interlayer voltage.

In Sec. \ref{sec2} starting from the microscopic Hamiltonian and
using the BCS-like many-particle wave function  we derive the
stationary continuity equation. In Sec. \ref{sec3} stationary
vortex states are considered and the problem of the lower critical
current is addressed. In Sec. \ref{sec4} the state with moving
vortices is investigated and the power of losses caused by such a
motion is computed. We find the power of losses is proportional to
the square of the tunneling amplitude and depends nonlinearly on
the resistance of the load.

\section{The model}
\label{sec2}

Let us consider a bilayer electron system subjected by a strong
perpendicular to the layers magnetic field $B$. The filling
factors of the layers satisfy the condition $\nu_1+\nu_2=1$, where
$\nu_i=2\pi\ell^2 n_i$, $n_i$, are the electron densities in the
layers,  and $\ell=\sqrt{\hbar c/e B}$ is the magnetic length.
Implying the energy gap between the Landau levels $\hbar\omega_c$
($\omega_c$ is the cyclotron frequency) be larger than the Coulomb
energy $e^2/\varepsilon\ell$ ($\varepsilon$ is the dielectric
constant) we use the lowest Landau level approximation. In this
approximation the Hamiltonian of the system has the form
\begin{equation}\label{1}
 \fl  H=-t\sum_X\left(a^+_{1X}a_{2X}+{\textrm H.c.}\right)+\frac{1}{2S}\sum_{n,n'=1,2}\sum_{\bf
  q}V_{n, n'}({\bf q})[\rho_n({\bf q})\rho_{n'}(-{\bf
  q})-\delta_{nn'}\rme^{-\frac{q^2\ell^2}{2}}\rho_n(0)].
\end{equation}
Here $a^+_{nX}$, $a_{nX}$ are the creation and annihilation
operators for the electrons in the $n$-th layer, $X$ is the
guiding center of the orbit, $S$ is the area of the layer, $t$ is
the interlayer tunneling amplitude,
 $\rho_n({\bf
 q})=\sum_X a^+_{n,X+q_y\ell^2/2}a_{n,X-q_y\ell^2/2}\rme^{\rmi q_x
 X-q^2\ell^2/4}$
is the operator for the Fourier component of the electron density,
and
 $V_{n,n'}({\bf q})=({2\pi e^2}/{\varepsilon
  q}) \exp(-q d|n-n'|)$
is the Fourier component of the Coulomb potential ($d$ is the
distance between the layers). In what follows we consider the
bilayer of a rectangle shape ($S=L_x \times L_y$) with the planar
currents  directed parallel to the $x$ axis.

The state with the interlayer phase coherence is described by the
wave function
\begin{equation}\label{4-0}
  |\Psi\rangle=\prod_X\left(u a_{1X}^+
  + v a_{2X}^+ \right)|0\rangle ,
\end{equation}
where the coefficients $u$ and $v$ satisfy the condition
$|u|^2+|v|^2=1$. The physical meaning of this function  and its
applicability for a description of the superfluid state of $e-h$
pairs in quantum Hall bilayers was discussed in \cite{t1,t2,t3}
and in a number of further papers. Here we just remind the main
points. The function (\ref{4-0}) can be presented in another
equivalent form
\begin{equation}\label{4}
  |\Psi\rangle=\prod_X\left(u
  + v a_{2X}^+ h^+_{1X}\right)|vac\rangle ,
\end{equation}
where $h^+_n=a_n$ is the creation operator for the hole, and the
vacuum state $|vac\rangle$ is the state in which  layer 1 is fully
occupied and the layer 2 is empty ($\nu_1=1$ and $\nu_2=0$).
 It was shown \cite{rs} that the function (\ref{4-0}), (\ref{4}) corresponds to the ground
state of the Hamiltonian (\ref{1}) in the limit $d/\ell \to 0$.
One can see that the function (\ref{4}) is the standard BCS form
for the exciton condensate. The exciton is thought of as an
electron in the layer 2 bound to a hole in the layer 1. One can
show (see, for instance, \cite{sim}) that in the first quantizied
language the function (\ref{4-0}),(\ref{4}) is reduced to the
(111) Halperin wave function. At finite $d/\ell$ the state
(\ref{4-0}), (\ref{4}) is not the exact ground state.
Nevertheless, numerical studies show (see e.g. Ref. \cite{sim1})
that at $d/l<0.5$ the overlap with the exact ground state is close
to hundred percents. For the system with given filling factors of
the layers the $u-v$ coefficients in (\ref{4-0}), (\ref{4}) read
as $u=\sqrt{\nu_1}$ and $v=e^{i \varphi}\sqrt{\nu_2}$. In the
absence of the interlayer tunneling the energy of the state
(\ref{4-0}), (\ref{4}) does not depend on the phase $\varphi$. In
case of spatially dependent phase $\varphi=\varphi(X)$ the
function (\ref{4-0}), (\ref{4}) describes the state with nonzero
counterflow electrical currents in the layers (see \cite{t1}).

The order parameter for the exciton condensate at $T=0$ is
\begin{equation}\label{5}
\Delta(X)=\langle\Psi| a_{2 X} h_{1 X}|\Psi\rangle= u^*_X
v_X=\Delta_0 e^{i\varphi(X)},
\end{equation}
where $\Delta_0=\sqrt{\nu_1(1- \nu_1)}$. Due to the absence of the
kinetic energy of carriers in the Landau level the
self-consistence equation for the order parameter at $T=0$ is
reduced to an algebraic equation. At nonzero temperature the
equation for the order parameter takes a self-consistent form
(\cite{kh, ml}) from which the temperature dependence of the order
parameter and the mean-field critical temperature can be obtained.
But due to the two-dimensional nature of the bilayer excitons the
temperature of transition into the superfluid state is not the
mean-field critical temperature,  but the
Berezinskii-Kosterlitz-Thouless transition temperature $T_{\rm
BKT}\approx \pi \rho_s/2$, see \cite{t1} (the definition of the
superfluid stiffness $\rho_s$ is given below).

If the gradient of the phase is small in comparison with the
inverse magnetic length, the energy of the system $E=\langle
\Psi|H|\Psi\rangle$ can be written in the continuous approximation
\begin{equation}\label{7}
  E=E_0+\int \rmd^2 r \left[ \frac{1}{2} \rho_{\rm s}  \left( \frac{\rmd
  \varphi}{\rmd x}\right)^2-\frac{\tilde{t}}{2\pi \ell^2} (\cos
  \varphi-1)\right],
\end{equation}
where $E_0$ is the energy of the ground state, $\tilde{t}=2 t
\Delta_0$ is the tunneling energy and
\begin{eqnarray}\label{8}
  \rho_{\rm s}=
  \Delta_0^2 \frac{e^2}{4 \pi\varepsilon \ell}
  \left[\sqrt{\frac{\pi}{2}} \exp\left(\frac{d^2}{2\ell^2}\right) {\rm
  erfc}\left(\frac{d}{\ell\sqrt{2}}\right)\left(1+\frac{d^2}{\ell^2}\right)
  -\frac{d}{\ell}\right]
\end{eqnarray}
is the energy parameter called the superfluid stiffness. The
superfluid stiffness can also be presented in a more familiar form
$\rho_{\rm s}=\hbar^2 n_{\rm s}/m_B$, where $n_{\rm s}$ is the
superfluid density of the pairs, and $m_B$ is the magnetic mass of
the pair (see, for instance, \cite{lr}). One can see from
(\ref{7}) that at nonzero tunneling amplitude the ground state
corresponds to the phase $\varphi=0$. But it does not mean the
fixation of the phase and the absence of electrical currents. It
means that the planar current should be accompanied with the
interlayer current.

Varying the energy (\ref{7}) with respect to the phase and
equating the result to zero we obtain the equation
\begin{equation}\label{8a}
- \frac{e \rho_{\rm s}}{\hbar}\frac{\rmd^2 \varphi}{\rmd
x^2}+\frac{e}{\hbar}\frac{\tilde{t}}{2\pi \ell^2} \sin \varphi=0.
\end{equation}
The first term in Eq. (\ref{8a}) is  the 2D divergence of the
density of the planar supercurrent
\begin{equation}\label{9}
 j_{1}=-j_{2}=-\frac{e \rho_{\rm s}}{\hbar} \frac{\rmd \varphi}{\rmd
 x}.
\end{equation}
The expression for the planar current can be obtained from the
gauge invariance arguments (see \cite{18,k1}).

One can see the Eq. (\ref{8a}) is the stationary continuity
equation
\begin{equation}\label{11}
\frac{\rm d j_{1}}{\rm d x}+I_{1\to 2}=0,
\end{equation}
where
\begin{equation}\label{10}
   I_{1\to 2}=\frac{e}{\hbar}\frac{\tilde{t}}{2\pi \ell^2} \sin
   \varphi
\end{equation}
is the density of the interlayer supercurrent flowing from the
layer 1 to the layer 2.

The coherent electrical current between the layers corresponds to
nonzero $\varphi$, in a close analogy with the Josephson effect
between two bulk superconductors. The difference is that $\varphi$
is the phase of a given condensate, but not the phase difference
for two condensates.

Eq. (\ref{8a}) can be presented in the form
\begin{equation}\label{12}
  \frac{\rmd^2 \varphi}{\rmd x^2}=\frac{1}{\lambda^2} \sin \varphi,
\end{equation}
where $\lambda=\ell\sqrt{2\pi\rho_{\rm s}/\tilde{t}}$ is the
Josephson length.
 As follows from Eq.
(\ref{12}),  the gradient of the phase is of order of
$\lambda^{-1}$. Therefore, the continuity approximation (\ref{7})
requires $\lambda\gg \ell$. This inequality is fulfilled if the
tunneling amplitude $t$ is much smaller than the Coulomb energy
$e^2/\varepsilon\ell$ and the filling factor $\nu$ is not very
close to zero.

\section{Stationary vortex state and lower input critical current}

\label{sec3}

The problem of the lower critical current was already discussed in
a number of papers \cite{sh1,18,19}. Here we revisit this question
again with the aim to give a more accurate definition of the lower
critical current and to clarify some discrepancies in
\cite{sh1,18,19}.

Let us consider the following situation. A fixed input current
from a source is entered into one layer in a given (source) end of
the system. There is no difference between the electrochemical
potentials of the layers and the stationary state is realized. The
current withdrawn from the adjacent layer at the same end is equal
to the input current. The values of the output and input currents
at the opposite (load) end are also equal to each other, but, in
general, they may differ from currents at the source end. For
instance, such a situation corresponds to the load geometry
\cite{j2} (then all input and output currents are equal to one
another). In the loop geometry the stationary state can be
realized if the load has zero resistance (superconducting load),
or if an additional source in the load circuit provides zero
difference of electrochemical potentials between the layers.

Depending on the values of the input currents two qualitative
different stationary current patterns can emerge: either the
planar and interlayer currents are nonzero in the whole bilayer,
or the currents decrease exponentially with the distance from the
ends and there is no currents in the internal part of the bilayer.
We define the lower critical input current as the current at which
the switching between these two regimes occurs. If the input
currents at both ends are lower than the critical one the second
pattern is realized and the interior part of the bilayer is not
involved into the transmission of the current (the cutting of the
bilayer does not change currents in outer circuits). If one or
both input currents exceed $j_c$ the first pattern emerges.

Let us switch to the quantitative analysis.  Eq. (\ref{12})
coincides in form with the equation of motion for a nonlinear
pendulum
\begin{equation}\label{13}
 \frac{\rmd^2 \varphi}{\rmd t^2}=\omega_0^2 \sin
\varphi,
\end{equation}
where $\omega_0$ is the resonant frequency of the pendulum, and
$\varphi$ is the angle coordinate counted from the unstable
equilibrium point. The formal coincidence of Eq. (\ref{12}) and
Eq. (\ref{13}) allows us to describe the stationary states in the
bilayer basing on known behavior of a nonlinear pendulum.

Depending on the energy, the pendulum oscillates, completes a full
revolution in infinite time, or rotates. The maximum angular
velocity $\omega_{\rm m}$ of the pendulum (velocity at
$\varphi=\pi$) is proportional to the square root of the energy
and it increases if one switches from the oscillating to the
rotating regime. For the full revolution in infinite time the
maximum angular velocity $\omega_{\rm m} =2\omega_0$.

The time dependent angular velocity in the pendulum problem
corresponds to the space dependent planar supercurrent in the
bilayer problem. The counterpart of the full revolution regime is
a state with one Josephson vortex. In this state the planar and
the interlayer supercurrents are given by the expressions
\begin{eqnarray}\label{14}
     j_{1}=-j_{2}=   j_{\rm c} {\rm sech}\left(
  \frac{x-x_0}{\lambda}\right),\cr
  I_{1\to 2}=  \frac{j_{\rm c}  }{\lambda} \, {\rm sech}\left(\frac{x-
  x_0}{\lambda}\right)
  \tanh \left(\frac{x-x_0}{\lambda}\right),
\end{eqnarray}
where $x_0$ is the vortex center, and
\begin{equation}\label{13a}
    j_{\rm c}=\frac{e\rho_{\rm s}}{\hbar}\frac{2}{\lambda}
\end{equation}
is the maximum value of planar supercurrent in the single vortex
state (it is reached at the center of the vortex). Note that Eq.
(\ref{13a}) can be obtained directly  from Eq. (\ref{9}) under
accounting the correspondences $(\rmd \varphi/\rmd x)_{\rm
max}\Leftrightarrow \omega_{\rm m}=2\omega_0$ and $\omega_0
\Leftrightarrow 1/\lambda$.

The counterpart of the rotation regime is the state with many
equally distanced Josephson vortices with the same sign of
vorticity. The supercurrents in this state read as
\begin{eqnarray}\label{16}
  j_{1}=-j_{2}=  \frac{j_{\rm c}}{\sqrt{\eta}} {\rm dn}\left( \frac{x}{
  \lambda\sqrt{\eta}},\eta\right), \cr
  I_{1\to 2}=  \frac{j_{\rm c}}{\lambda} \, {\rm
  sn}\left( \frac{x}{\lambda\sqrt{\eta}},\eta\right)
  {\rm cn}\left( \frac{x}{\lambda\sqrt{\eta}},
  \eta\right).
\end{eqnarray}

The analog of the oscillating motion of the pendulum is the
multivortex state with vortices of alternating vorticity:
\begin{eqnarray}\label{16a}
  j_{1}=-j_{2}=  {j_{\rm c}}{\sqrt{\eta}} {\rm cn}\left( \frac{x}{
  \lambda},\eta\right), \cr
  I_{1\to 2}=  \frac{j_{\rm c} \sqrt{\eta}}{\lambda}  {\rm
  dn}\left( \frac{x}{\lambda},\eta\right)
  {\rm sn}\left( \frac{x}{\lambda},
  \eta\right).
\end{eqnarray}
In Eqs. (\ref{16}), (\ref{16a})  ${\rm dn}(u, \eta)$, ${\rm cn}(u,
\eta)$ and ${\rm sn}(u, \eta)$ are the Jacobi elliptic functions
and the parameter $\eta$ belongs to the interval $(0,1)$.

The boundary condition for the bilayer problem is the condition on
the gradients of the phase. This condition alone does not
determine an unique current state. The system chooses the state
that at given input currents has the lowest energy. Direct
calculation shows that the energies of  the multivortex state
(\ref{16}) and (\ref{16a}) are both higher than the energy of the
single vortex state (\ref{14}). It is obvious result since the
energy of any multivortex state is proportional to the number of
vortices.

Here we should emphasize the difference between the pendulum
problem and the bilayer problem. The energy of the pendulum is the
integral of motion of Eq. (\ref{13}), while the bilayer energy
(\ref{7}) is not the first integral of Eq. (\ref{12}). Therefore,
the analogy between these two problems fails if one analyzes the
energies of different states.

One can see that  the planar supercurrent may exceed $j_{\rm c}$
only in the state (\ref{16}). Therefore, if the input current
$j_{\rm in}$ is higher than $j_{\rm c}$ the multivortex state
(\ref{16}) is realized. The parameter $\eta$ is determined by the
conditional minimum of energy and it is equal to $\eta=(j_{\rm
c}/j_{\rm in})^2$. The density of the Josephson vortices increases
under increase of the input current.

If the input current is smaller than $j_{\rm c}$ the state with
one incomplete vortex centered outside the bilayer satisfies the
boundary condition and corresponds to the minimum of energy. Thus,
the state realized at $j_{\rm in}\leq j_{\rm c}$ is the state with
zero supercurrents inside the bilayer.

The vortex state with alternating vorticities (\ref{16a}) is not
realized in the bilayers. It may satisfy the boundary conditions
only at $j_{\rm in}< j_{\rm c}$, but its energy is larger than the
energy of the state with one vortex. It contradicts the conclusion
of Ref. \cite{19}, but in Ref. \cite{19} the first integral of Eq.
(\ref{12}) was incorrectly identified with the energy.

Thus, the quantity (\ref{13a}) is just the lower input critical
current defined above.

At $j_{\rm in}\gg j_{\rm c}$ the overlapping between the vortices
is large and the current pattern is approximated by a sum of large
d.c part and small harmonic a.c. part:
\begin{eqnarray}\label{16b}
\varphi\approx\pi- kx- A_\varphi
 \sin kx , \cr
  j_{1}=-j_{2}\approx  \frac{j_{\rm in}}{1+A_\varphi}
  \left(1 +A_\varphi\cos kx\right),\cr
  I_{1\to 2}\approx  \frac{e}{\hbar}\frac{\tilde{t}}{2\pi \ell^2} \sin
  kx,
  \end{eqnarray}
where $$
    k=\frac{\pi j_{\rm in}}{j_{\rm c} \lambda
    K\left(\frac{j_{c}^2}{j_{\rm in}^2}\right)}\approx \frac{2 j_{\rm in}}{j_{\rm c}
    \lambda}, \quad A_\varphi=\frac{j_{\rm c}^2}{4 j_{\rm in}^2}.
$$ Here $K(\eta)$ is the complete elliptic integral of the first
kind.

To conclude this section we note that at small input current
(lower than $j_{\rm c}$) the tunneling provides shunting of the
bilayer at both ends. It means that the load geometry  experiments
proposed in \cite{j2} for the observation of e-h superfluidity
should be done at large currents.

\section{Vortex motion and dissipation}

\label{sec4}

As was already discussed in the introduction, the stationary state
cannot be realized in the bilayer transmission line  with nonzero
load resistance.  But even in case of non-stationary currents
inside a bilayer the currents in external circuits may remain
stationary one. The latter occurs if the contacts between the
layers and the external wires average the input and output
currents.

In such a situation  the term "lower critical current" slightly
changes its meaning. The average current depends on the density of
vortices and can be smaller than $j_{\rm c}$. But the input
current higher than critical one is required at the beginning of
the process to induce a chain of moving vortices.

To study the non-stationary vortex state we will use the set of
equations for the phase of the order parameter $\varphi$ and for
the local difference of the electrochemical potentials between the
layers $eV$, where $V$ is the local interlayer voltage. The first
equation comes from the non-stationary continuity equation
\begin{equation}\label{17b}
e\frac{\partial \tilde{n}}{\partial t}+\frac{\partial j_1^{\rm
s}}{\partial x}+ I_{1\to 2}+\frac{\partial j_{1}^{\rm n}}{\partial
x}=0.
\end{equation}
Here $\tilde{n}$ is the excess local density of electron-hole
pairs. It is connected with the local voltage by the capacitor
equation
\begin{equation}\label{19}
    \frac{e \tilde{n}}{V}=C,
\end{equation}
where
\begin{equation}\label{20}
    C=\frac{\varepsilon}{4\pi d} \frac{1}{1-\frac{\ell}{d}\sqrt{\frac{\pi}{2}}
    \left[1-\exp\left(\frac{d^2}{2\ell^2}\right){\rm erfc}\left(
    \frac{d}{\ell \sqrt{2}}\right)\right]}
\end{equation}
is the effective capacity of the bilayer system per unit area. The
effective capacity takes into account the exchange interaction
between the layers and differs from the classical capacity by the
additional factor (the second factor in Eq. (\ref{20})). It can be
obtained from the dependence of the energy on the filling factor
imbalance (see, e.g. \cite{ml}). The superscripts "s" and "n" in
Eq. (\ref{17b}) stand for the planar supercurrent and for the
planar current of quasiparticles. The current of qusiparticles is
taken into account in Eq. (\ref{17b}) because it is induced by
a.c. planar electrical fields that emerge in the non-stationary
state. These fields equal to $E_1=-E_2=-(1/2)\partial V/\partial
x$ and the quasiparticle current can be presented in the form
$j_1^{(n)}=-(\sigma_n/2)\partial V/\partial x$, where $\sigma_n$
is the quasiparticle conductivity. We do not take into account the
normal component of the interlayer current because the normal
tunneling is suppressed due to discreteness of the Landau levels.

The second equation on $\varphi$  and $V$ can be derived from the
equations of superfluid hydrodynamics \cite{21}, namely, from the
equation for the superfluid velocity with dissipative terms. The
linearized version of this equation reads as
\begin{equation}\label{18a}
     \frac{\partial {\bf v}_{\rm s}}{\partial t} = - \nabla \mu_m
    +\nabla\left[\zeta_3\nabla\cdot n_{\rm s}({\bf v}_{\rm s}-{\bf v}_{\rm n})+\zeta_4
    \nabla\cdot {\bf v}_{\rm n}\right],
\end{equation}
where $\mu_m$ is the chemical potential per unit mass
$\mu_m=\mu/m$ ($m$ is the mass of the Bose particle), $n_{\rm s}$
is the superfluid density, ${\bf v}_{\rm n}$ is the velocity of
the normal component, $\zeta_3$, $\zeta_4$ are the second
viscosity parameters. To apply Eq. (\ref{18a}) for  the
description of the bilayer electron-hole superfluidity  one should
replace the divergence of the superfluid flow  in (\ref{18a}) with
the divergence of the planar supercurrent plus the tunnel
supercurrent: $\nabla(n_{\rm s} {\bf v}_{\rm s}) \Rightarrow
(1/e)(\partial j_{1}^{\rm s}/\partial x + I_{1\to 2})$, and
substitute the difference of the electrochemical potentials
instead of the chemical potential $\mu\Rightarrow eV$. Using Eqs.
(\ref{10}), (\ref{9}), and the expression for the superfluid
velocity ${\bf v}_{\rm s}=-\hbar \nabla\varphi/m$ (the negative
sign is because the phase $\varphi$ describes the coherence
between the electrons), we obtain
\begin{equation}\label{18b}
     \nabla \left(\hbar \frac{\partial {\varphi}}{\partial t}\right) = \nabla \left[e
     V  +\zeta_3 \frac{m}{2 \pi\ell^2 \hbar}
    \left(2\pi \ell^2 \rho_{\rm s} \frac{\partial^2
    \varphi}{\partial x^2}-\tilde{t} \sin \varphi\right)
    +\Or(v_{\rm n}) \right].
\end{equation}
 The quantity $\Or(v_{\rm n})$ in Eq. (\ref{18b})
stands for the term caused by the motion of the normal component.
Below we will neglect the $\Or(v_{\rm n})$ term. One can show that
this term yields the correction to the power of losses quadratic
in the dissipative coefficients, while  the leading term is linear
in these coefficients. Since we are interested in the vortex
solution, the gradients in (\ref{18b}) can be omitted.

Thus,  we obtain  the following set of equations for $\varphi$ and
$V$
\begin{equation}\label{18}
     C\frac{\partial V}{\partial t}= \frac{e}{\hbar}\rho_{\rm s} \frac{\partial^2
    \varphi}{\partial x^2} -\frac{e}{\hbar}\frac{\tilde{t}}{2\pi \ell^2} \sin
    \varphi + \frac{\sigma_{\rm n}}{2} \frac {\partial^2
    V}{\partial x^2},
\end{equation}
\begin{equation}\label{21}
     \hbar \frac{\partial \varphi}{\partial t}=e
     V+\alpha_\varphi\left(2\pi \ell^2 \rho_{\rm s} \frac{\partial^2
    \varphi}{\partial x^2}-\tilde{t} \sin \varphi\right).
\end{equation}
 The coefficient
$\alpha_\varphi=\zeta_3 m/2 \pi\ell^2 \hbar$ in Eq. (\ref{21}) is
the second superfluid viscosity expressed in dimensionless units
(with $m=m_B$, the magnetic mass of the e-h pair). Eqs.
(\ref{18}), (\ref{21}) contain two terms responsible for the
dissipation. The term proportional to $\sigma_{\rm n}$ describes
the Joule losses caused by a.c. quasiparticle currents, and the
term proportional $\alpha_\varphi$ - the losses caused by the
second viscosity.  The equations with the same form of dissipative
terms were used in \cite{22}.  Eq. (\ref{21}) was also obtained in
Ref. \cite{23} for the electron-hole bilayers in zero magnetic
field in the "dirty" limit.

The channels of dissipation caused by the second viscosity and
quasiparticle currents are present in any superfluid system, but
they do not result automatically in dissipative losses. One can
see that in the stationary regime the dissipative terms in Eqs.
(\ref{18}), (\ref{21}) equal to zero. The losses appear under
presence of additional factors. For the bilayer transmission line
the main factor is non-stationarity of the interlayer and the
planar supercurrents.

As was shown in the previous section, the input current may induce
Josephson vortices in the bilayer system. At nonzero difference of
the electrochemical potentials the vortices begin to move from the
source to the load. Neglecting for a moment the dissipative terms
we obtain from Eqs. (\ref{18}), (\ref{21}) the following equation
for the phase
\begin{equation}\label{24}
\frac{\partial^2 \varphi}{\partial \xi^2}-\frac{\partial^2
\varphi}{\partial \tau^2}=\sin \varphi,
\end{equation}
 where the dimensionless variables
 $\xi=x/\lambda$ and
$\tau=t/\tau_0$ ($\tau_0=(\hbar\ell/e)\sqrt{2\pi C/\tilde{t}}$)
are used. Eq. (\ref{24}) has the solution
$\varphi(\xi,\tau)=\phi(\xi-\beta \tau)$, where $\beta\ne 1$ and
$\phi(u)$ satisfies the equation
\begin{equation}\label{25}
\frac{\partial^2 \phi}{\partial u^2}=\frac{1}{1-\beta^2}\sin \phi
\end{equation}
that up to the notations coincides with Eq. (\ref{12}). Using the
results of Sec. \ref{sec3} we obtain the following expression for
the planar current and for the interlayer voltage
\begin{equation}\label{27}
    j_{1}^{\rm s}(x,t)=  \frac{2 e}{\hbar} \rho_{\rm s} k\
    {\rm dn}
    \left(k x- \omega t, \eta \right),
\end{equation}
\begin{equation}\label{29}
    V(x,t)=\frac{2 \hbar \omega}{e}{\rm dn}
    \left(k x- \omega t, \eta \right),
\end{equation}
where
\begin{equation}\label{28}
    k=\frac{1}{\lambda\sqrt{\eta|1-\beta^2|}},
    \quad \omega=\frac{\beta}{\tau_0\sqrt{\eta|1-\beta^2|}}.
\end{equation}

Eq. (\ref{27}) describes the moving vortex lattice. The velocity
of motion is
\begin{equation}\label{33}
    u_{\rm v}=\frac{\omega}{k}=\frac{e^2}{\hbar^2}\rho_{\rm s} \frac{\langle
    V\rangle}{\langle j_{1}^{\rm s}\rangle}
    =\frac{e^2 \rho_{\rm s} R  L_y}{\hbar^2}.
\end{equation}
Here and below $\langle\ldots\rangle$ means the time average. In
Eq. (\ref{33}) $R$ is the resistance of the load (we imply the Ohm
law $\langle
    V\rangle=R L_y \langle j_{1}^{\rm s}\rangle$ for the load circuit).
One can see that the vortex velocity is proportional to the load
resistance and does not depend on the input current. The parameter
$\beta=u_v \tau_0/\lambda$ is also proportional to the load
resistance.

Let us switch to the case of small but nonzero dissipation. The
power of losses can be obtained from the common expression for the
Joule losses
\begin{equation}\label{40}
    \Delta P=L_y\left\langle\int_0^{L_x} d x \left[I_{1\to 2} V
    + (j_{1}^{\rm s}+j_{1}^{\rm n})
    \left(-\frac{\partial V}{\partial
    x}\right)\right]\right\rangle.
\end{equation}
Integrating (\ref{40}) by parts and taking into account the
continuity equation (\ref{17b}) we obtain the obvious result that
the power of losses is the input power minus the output power:
\begin{equation}\label{43}
    \Delta P= (j_{\rm in}- j_{\rm out})V_0 L_y.
\end{equation}
Here $j_{out}$ is the output planar current, and $V_0$ is the
voltage in the output circuit (that coincides with the voltage
applied to the bilayer system at the input end). In deriving
(\ref{43}) we take into account that $\langle V (\partial V/
\partial t)\rangle=0$ for any function $V(x,t)$ periodic in $t$.

The difference between the input and the output average currents
emerges if the average value of the interlayer current
$\bar{I}=\langle I_{1\to 2} \rangle$ differs from zero. In what
follows we consider the case where the power of losses is much
smaller than the input power. In this case
 one can neglect the dependence of $\bar{I}$
on $x$ and approximate the difference $j_{\rm in}-j_{\rm out}$ as
$j_{\rm in}-j_{\rm out}=L_x \bar{I}$. Then the power of losses
reads as
\begin{equation}\label{43a}
   \Delta P= \bar{I} V_0 S.
\end{equation}
The quantity $\bar{I}$ can be found from the solutions of Eqs.
(\ref{18}), (\ref{21}). Here  we consider  the case of large input
current $j_{\rm in}\gg j_{\rm c}$. In this case one can seek for a
solution of Eqs. (\ref{18}), (\ref{21}) in the following form
\begin{eqnarray}\label{34}
    \varphi(x,t)=\omega t - k x + A_\varphi \sin(\omega t - k
    x)+B_\varphi \cos(\omega t - k x), \\
    \label{34a}
V(x,t)=V_0\left(1+A_V \sin(\omega t - k
    x)+B_V \cos(\omega t - k x)\right),
\end{eqnarray}
where $\omega=e V_0/\hbar$, and the quantity $k$ is connected with
the planar supercurrent by the relation $\langle j_1^{\rm
s}\rangle=e\rho_{\rm s} k/\hbar$. Substituting Eqs. (\ref{34}),
(\ref{34a})  into Eqs. (\ref{18}) (\ref{21}) and neglecting the
terms quadratic in $\tilde{t}$ we obtain the coefficients
$A_\varphi$, $B_\varphi$, $A_V$, $B_V$ in the leading order in
$\tilde{t}$:
\begin{eqnarray}
\label{35-3} \fl A_\varphi=-\frac{\tilde{t}}{2\pi \ell^2 k^2
\rho_{\rm
s}}\frac{(1-\beta^2+2\pi\ell^2k^2\alpha_\varphi\tilde{\sigma})
(1+2\pi\ell^2k^2\alpha_\varphi\tilde{\sigma})
+\alpha_\varphi\frac{V_0}{V_C}\left(\alpha_\varphi\frac{V_0}{V_C}
+\tilde{\sigma}\frac{V_0}{V_\rho}\right)}{D},\\
\label{35-4} B_\varphi=\frac{\tilde{t}}{2\pi \ell^2 k^2 \rho_{\rm
s}}\frac{\tilde{\sigma}\frac{V_0}{V_\rho}\left(1
+2\pi\ell^2k^2\alpha_\varphi\tilde{\sigma}\right)+\alpha_\varphi\frac{V_0}{V_C}
\beta^2}{D},\\
\label{35} A_V=-\frac{\tilde{t}}{2\pi \ell^2 k^2
\rho_{\rm s}}\frac{\alpha_\varphi\frac{V_0}{V_C}+\tilde{\sigma}\frac{V_0}{V_\rho}}{D},\\
\label{35-2}
 B_V=-\frac{\tilde{t}}{2\pi \ell^2 k^2
\rho_{\rm s}}\frac{1+2\pi\ell^2
k^2\alpha_\varphi\tilde{\sigma}-\beta^2}{D},
\end{eqnarray}
where
\begin{equation}\label{39}
    D=\left(1-\beta^2+2\pi\ell^2 k^2 \alpha_\varphi\tilde{\sigma}\right)^2+
\left(\alpha_\varphi\frac{V_0}{V_C}+\tilde{\sigma}\frac{V_0}{V_\rho}\right)^2,
\end{equation}
and the following notations are used
\begin{equation}\label{36}
    \beta^2=\frac{C V_0^2}{\rho_{\rm s} k^2}, \quad
    V_C=\frac{e}{2\pi\ell^2 C}, \quad V_\rho=\frac{\rho_{\rm s}}{e},
    \quad
    \tilde{\sigma}=\frac{\sigma_{\rm n} \hbar}{2 e^2}.
\end{equation}
Note the equivalence of $\beta$ in (\ref{36}) and the definition
of $\beta$ given above.

 One can see that at
$\tilde{t}=0$ all the coefficients $A_\varphi$, $B_\varphi$,
$A_V$, $B_V$ are equal to zero that reflects the absence of
vortices in the bilayers with zero interlayer tunneling.

Using Eqs. (\ref{10}) and (\ref{34}) one finds the average value
of the interlayer current
\begin{equation}\label{41}
    \bar{I}=\frac{e\tilde{t}}{2\pi\ell^2
    \hbar}\langle \sin \varphi \rangle\approx
    \frac{e \tilde{t}}{4\pi \ell^2 \hbar} B_\varphi.
\end{equation}

Since $B_\varphi$ is proportional to $\tilde{t}$
 the average interlayer current is proportional to the square of
the matrix element of the interlayer tunneling. The higher order
corrections to $B_\varphi$ yield the contributions $\propto
\tilde{t}^3$ into $\bar{I}$. Therefore, for obtaining the power of
losses in the leading order it is enough to use the linear in
$\tilde{t}$ approximation for the phase.

Using Eqs.(\ref{41}) and (\ref{35-4}) we obtain the following
expression for the power of losses
\begin{equation}\label{45-0}
 \frac{\Delta P}{S}=\frac{\tilde{t}^2}{4\pi\ell^2 \hbar} F\left(\frac{R}{R_0}\right)
\end{equation}
where
\begin{equation}\label{45}
\fl F\left(\frac{R}{R_0}\right)=  \frac{\tilde{\sigma}
    \frac{V_C}{V_\rho} \left(\frac{R}{R_0}\right)^2 +\alpha_\varphi
    \left(\frac{R}{R_0}\right)^4}{\left(1-
    \left(\frac{R}{R_0}\right)^2
    +2\pi\alpha_\varphi\tilde{\sigma}\left(\frac{\bar{j}}{j_0}\right)^2\right)^2+
    2\pi \left(\frac{R}{R_0}\right)^2 \left(\frac{\bar{j}}{j_0}\right)^2  \frac{V_\rho}{V_C}
    \left( \alpha_\varphi +\tilde{\sigma}\frac{V_C}{V_\rho}\right)^2}.
\end{equation}
 Here $j_0=e \rho_{\rm s}/\hbar\ell$ ($j_0\gg j_{\rm c}$) is
the current of order of the higher critical current \cite{18} (the
current above which superfluid state is destroyed). The quantity
$R_0=\hbar/eL_y\sqrt{\rho_{\rm s} C}$ is the resonant load
resistance. One can see that dissipative losses increases
considerable at $R$ approaches to $R_0$ (the condition $R=R_0$ is
equivalent to $\beta=1$). We emphasize that the approximation
solution (\ref{34}) and the answer (\ref{45}) are not valid in a
resonant case  and at $R\approx R_0$ they describe the situation
only qualitatively.

Under obtaining Eq. (\ref{43a}) we take into account the relation
$\langle j_{1}^{\rm s}\rangle=\bar{j}=\frac{e}{\hbar}\rho_{\rm s}
k$ and the  Ohm's law $V_0\approx R_l L_y \bar{j}$ in the load
circuit. In Eqs.(\ref{45}) the terms in nominator quadratic in
dissipative coefficients are omitted.

Thus we conclude that the power of losses is proportional to the
square of the tunneling amplitude and depends nonlinearly on the
load resistance.  At small load resistance $R\ll R_0$ the
dissipation is connected in the main part with the conductivity of
quasiparticles. It is proportional to the square of the load
resistance
\begin{equation}\label{46}
   \frac{\Delta P}{S}\approx\frac{\tilde{t}^2}{4\pi\ell^2 \hbar}
    \tilde{\sigma}\left(\frac{R}{R_0}\right)^2
    \frac{V_C}{V_\rho}
\end{equation}
and vanishes at  $R=0$ . At large load resistance $R\gg R_0$ the
main contribution the into the power of losses comes from the
second viscosity. The power of losses at large $R$ approaches to
the constant value
\begin{equation}\label{47}
   \frac{\Delta P}{S}\approx\frac{\tilde{t}^2}{4\pi\ell^2 \hbar} \alpha_\varphi.
\end{equation}
The resonant resistance $R_0$ does not depend on the tunneling
amplitude but it can be tuned by a change of $d/\ell$ (it is the
increase function of that parameter). At $d/\ell =1$ the resonant
resistance is approximated as
\begin{equation}\label{47-1}
 R_0\approx \frac{h}{e^2}\frac{2}{\Delta_0}\frac{\ell}{L_y}
\end{equation}
At low load resistance the "bottle neck" of the load circuit are
the contacts where the interlayer phase coherence is broken. We
evaluate from (\ref{47-1}) that at longitudinal resistivity of the
layer $\rho_{xx}=1$k$\Omega$ the resistance of the arm $R\sim R_0$
correspond to the length of contacts $\sim 10^2\ell\approx 1\mu$m.
Since it is rather small length we conclude that the experimental
situation  corresponds most probably to the case $R\gg R_0$.

In conclusion we note that in the non-resonant regime the power of
losses does not depend on the input current. Since the input power
is proportional to the input current, the efficiency factor
increases under increase of the input current.

\section{Conclusion}

\label{disc}

We have shown that in the bilayer system with superfluid
electron-hole pairs the dissipation appears under the transmission
of the current from the source to the load. The effect is
connected with that the electrical currents inside the bilayer
becomes non-stationary at nonzero difference of the
electrochemical potentials between the layers.

The non-stationary state can be interpreted as the state with
moving Josephson vortices. It it well known that in type II
superconductors   the motion of quantum vortices results in
dissipation, but the latter can be eliminated by pinning of the
vortices. In this connection one can think that the pinning may
also suppress the dissipation in  the bilayers. But it is not
true. The difference between type II superconductors and the
bilayers is the following. In type II superconductors nonzero
voltage along the direction of the supercurrent emerges due to the
vortex motion. In the bilayers nonzero interlayer voltage is
required to support electrical current in the load circuit and
this voltage causes the motion of Josephson vortices. The state is
stationary only at zero interlayer voltage at which there is no
current in the load circuit.

 One can also arrive at this
conclusion in another way. The energy of a single vortex reads as
\begin{equation}\label{50}
    E_{\rm v}=\frac{4L_y}{\ell}\sqrt{\frac{2}{\pi}}\sqrt{\rho_{\rm s} \tilde{t}}
\end{equation}
According to Eq. (\ref{50}) the pinning may occur due to spatial
variation of the tunneling amplitude. In the latter case an
approximate solution for the phase can be found by the same way as
was  done in previous section. The only difference that the
coefficients $A_\varphi$, $B_\varphi$, in Eq. (\ref{34}) should be
replaced with $x$-dependent quantities. The solution Eq.
(\ref{34}) with spatially dependent amplitudes $A_\varphi$ and
$B_\varphi$ also corresponds to a non-stationary state. It differs
from the state described in the previous section by that the phase
remains non-stationary in any reference frame, including the frame
in which the vortices are at rest. One can show that in this case
the dissipative losses are also nonzero and proportional to
dissipative coefficients $\alpha_\varphi$ and $\tilde{\sigma}$.

Thus, nonzero interlayer tunneling  results in that the
electron-hole pairs in the bilayer cannot transmit electrical
energy without dissipation. Nevertheless, since the power of
losses is proportional to the square of the tunneling amplitude
and this amplitude depends exponentially on the interlayer
distance one can expect that it is possible to create bilayer
systems with negligible small dissipation.

\end{document}